# Scientific Computing Using Consumer Video-Gaming Hardware Devices

Mr. Glenn Volkema, Dr. Gaurav Khanna, *Center for Scientific Computation & Visualization Research.*

*Abstract*—Commodity video-gaming hardware (consoles, graphics cards, tablets, etc.) performance has been advancing at a rapid pace owing to strong consumer demand and stiff market competition. Gaming hardware devices are currently amongst the most powerful and cost-effective computational technologies available in quantity. In this article, we evaluate a sample of current generation video-gaming hardware devices for scientific computing and compare their performance with specialized supercomputing general purpose graphics processing units (GPGPUs). We use the OpenCL SHOC benchmark suite, which is a measure of the performance of compute hardware on various different scientific application kernels, and also a popular public distributed computing application, *Einstein@Home* in the field of gravitational physics for the purposes of this evaluation.

*Index Terms*—Scientific Computing, Accelerators, Parallel Computing, Supercomputing, GPU, GPGPU, OpenCL, SHOC, Physics

## I. INTRODUCTION

THERE is considerable current interest in harnessing the advancements made in multi- and many-core technology for scientific high-performance computing (HPC). An example of this trend is the rapid rise in the use of custom-designed HPC general purpose graphics processing units (GPGPUs) as "accelerators" in workstations and even large supercomputers. In fact, the second-fastest supercomputer today, ORNL's *Titan*, makes use of Nvidia's custom-HPC Tesla (Kepler series) GPUs to achieve *petascale* performance [1] and there are over 100 such accelerated systems in the top 500 supercomputers worldwide. One reason for this recent trend is that the large consumer video-gaming market significantly aids in "subsidizing" the research and development cost associated towards advancing these compute technologies, resulting in high performance at a low cost. In addition to their cost effectiveness, these GPU technologies

This article was submitted for peer-review on July 1st, 2016. This work was supported in part by the NSF award PHY-141440 and by the US Air Force agreement 10-RI-CRADA-09.

Mr. Glenn Volkema, is a member of the Physics Department and the Center for Scientific Computing and Visualization Research at the University of Massachusetts Dartmouth, North Dartmouth, MA 02747 USA (e-mail: gvolkema@umassd.edu).

Dr. Gaurav Khanna is a Professor in the Physics Department and the Associate Director of the Center for Scientific Computing and Visualization Research at the University of Massachusetts Dartmouth, North Dartmouth, MA 02747 USA (e-mail: gkhanna@umassd.edu).

are substantially "greener" over CPUs, delivering higher computational performance per Watt of electrical-power consumed [2].

In a similar spirit, there is also the opportunity to utilize the commodity "off-the-shelf" video-gaming hardware itself for scientific computing. Here we are specifically referring to consumer-grade video-gaming cards and also game consoles themselves, as opposed to the custom-HPC variants. These often have similar high-performance characteristics and in addition, are sold at a significant discount (sometimes even *below* manufacturing cost) because the business model allows for making up the deficit through the sales of video games and other application software. Starting in 2007 the authors were pioneers in successfully utilizing a cluster of *Sony PlayStation 3* (PS3) [3] gaming consoles for scientific computation [4]. They were able to demonstrate *order-of-magnitude* gains in efficiency metrics such as *performance-per-dollar* and *performance-per-Watt* as compared with traditional compute clusters [5,6,7]. Since then many universities and research groups took a similar approach and evaluated the value of such PS3 hardware for their own computation needs. The largest such system has been built by the *Air Force Research Laboratory* (AFRL) in Rome, NY. This system, named "AFRL CONDOR" utilizes 1,716 PS3s alongside traditional servers and Nvidia Tesla (Fermi series) GPGPUs to achieve 500 TFLOPS of computing power [8]. CONDOR has been demonstrated to be over 10× more cost-effective in similar metrics [9].

In this article we explore the capabilities of current generation consumer-grade, video-gaming hardware for scientific high-performance computing. Specific examples of the compute hardware that we consider interesting for this study are current gaming graphics cards like the AMD Radeon [10], Nvidia GeForce [11] series, and also the CPU-GPU "fused" or *heterogeneous* processor architectures like the AMD's Accelerated Processing Unit (APU) [12] and the Nvidia's Tegra System-on-a-Chip (SoC) [13]. The main advantage of considering such consumer-grade hardware for scientific HPC is low cost and high power-efficiency. Rapid advances and significant innovation is being enabled through major investments made by the gaming industry. This is driven by strong consumer demand for immersive gaming experiences that require computational power. High volume and intense competition keep costs low, while improvements to power-efficiency are forced by the engineering challenges and costs associated with dissipating increasing amounts of waste heat from a discrete device and the limited battery life of a mobile device.

The parallel software development framework that we will utilize in this work is the Open Computing Language (OpenCL) [14]. This is because OpenCL is a free, open standard and is vendor and platform neutral. It allows one to write high performing, yet, *portable* parallel code that executes on a wide variety of processor architectures including CPUs, GPUs, FPGAs, DSPs, SoCs and the Xeon Phi. All major processor vendors (Nvidia, AMD, Altera, TI, Qualcomm, ARM, IBM, Intel, etc.) have adopted the OpenCL standard and have released support for it on their hardware. The availability of the OpenCL parallel programming framework on such a wide variety of compute hardware, including consumer-grade gaming GPUs, allows for the possibility of easily evaluating such hardware for scientific computation.

This article is organized as follows: Section II provides a description of the metrics and the scientific application benchmarks we consider in this work; Section III includes detailed information on the hardware devices we evaluate; Section IV focuses on a discussion of the results we obtained and finally in Section V we end with some conclusive remarks. We also include a detailed Appendix that includes all the explicit data that our benchmarking generated.

## II. Scientific Applications

In this section, we present a brief introduction to the scientific applications we utilize to evaluate the overall performance of the gaming hardware under consideration in this work. We will not only study the overall performance delivered by the different gaming devices under consideration in this work, but also compare the associated costs involved through common efficiency metrics such as *performance-per-dollar* (a measure of the cost of procurement) and *performance-per-Watt* (operating costs). The latter is also an important current consideration in the context of "green" computing with the mission of not only making the research computing enterprise more eco-friendly, but also to enable the next generation of (*exascale*) supercomputing.

### A. SHOC Benchmark

The Scalable Heterogeneous Computing (SHOC) benchmark [15] is a scientific computing kernel suite based on OpenCL. It targets multi-core CPUs and many-core GPUs in a single or (message-passing based) clustered environment. For the purposes of this work, we only focus on a single device benchmark leaving the cluster-based study for future exploration. Below is the detailed list of compute kernels in the SHOC suite with a brief explanation:

Level 0 - Feeds & Speeds

• *Bus Speed Download and Readback*---This kernel measures the bandwidth of the interconnection bus between the host CPU and the GPU device.

• *Peak FLOPS*---This kernel measures the peak floating-point (single or double precision) operations per second.

• *Device Memory Bandwidth*---This kernel measures bandwidth for all types of GPU memory (*global, local, constant,* and *image*).

• *Kernel Compilation*---This benchmark measures average compilation speed and overheads for various OpenCL kernels.

Level 1 – Basic Parallel Algorithms

• *FFT*---This benchmark measures the performance of a 2D Fast Fourier Transform (FFT) for both single- and double-precision arithmetic.

• *GEMM*---This kernel measures the performance on a general matrix multiply BLAS routine with single- and double- precision floating-point data.

• *MD*---This kernel measures the speed of the Lennard-Jones potential computation from molecular dynamics (single- and double- precision tests are included).

• *Reduction*---This kernel measures the performance of a sum reduction operation using floating-point data.

• *Scan*---This kernel measures the performance of the parallel prefix sum algorithm on a large array of floating-point data.

• *Sort*---This kernel measures performance for a very fast radix sort algorithm that sorts key-value pairs of single precision floating point data.

• *SpMV*---This benchmark measures performance on sparse matrix with vector multiplication in the context of floating-point data, which is common in some scientific applications.

• *Stencil2D*---This kernel measures performance for a standard 2D nine-point stencil calculation.

• *Triad*---This benchmark measures sustainable memory bandwidth for a large vector dot product operation on single precision floating-point data.

Level 3 – Real Application Kernels

• *S3D*---This benchmark measures performance in the context of a simulation of a combustion process. It computes the rate of chemical reactions on a regular 3D grid.

### B. Einstein@Home

Laser Interferometer Gravitational Wave Observatory (LIGO) [16] is a National Science Foundation (NSF) funded facility dedicated to the experimental detection of gravitational radiation. It is part of an international network of detectors located across the globe (two LIGO sites in the United States, GEO/Virgo in Europe, TAMA/KAGRA in Japan and soon LIGO-India/IndIGO in India). Critical to a successful detection is the use of theoretical signal template banks from likely sources (like binary black hole systems) for the purpose of *matched-filtering* [17] (mainly due to the fact that the signal-to-noise ratio in the data streams is very low). LIGO has recently made the *first-ever* direct detections [18,19] of gravitational waves – "ripples in the fabric of spacetime" that Einstein predicted precisely 100 years ago as a consequence of his new theory of general relativity. These observatories generate data at the rate of several *petabytes per year* and they require highly computationally intensive data analysis [20]. *Einstein@Home* [21] is an open-source, NSF funded, public distributed computing project that offloads the big data analysis associated to these observatories to volunteers worldwide. The goal of *Einstein@Home* is finding gravitational waves emitted from neutron stars, by running a brute force search for different waveforms in an extremely large data-set. Thus, this project plays an important role in providing crucial data analysis capabilities to a very large and significant, international experiment and also deeply engaging the public in the relevant science. Since 2009 the project expanded its search for pulsar candidates to include radio-telescope data, and has had considerable success [22].

*Einstein@Home* involves over 430,000 participants with over 7 million client computers and has a total compute capability of over 2.2 PFLOPS. In addition to engaging citizens deeply into the science, the project has also discovered over 20 new pulsars through the radio-data search [23]. One of the authors has been a volunteer developer for the project since its inception in 2005 and has contributed to the development of the code on Apple PowerPC systems, PS3s and also GPUs [24,25]. With the major shift towards usage of low-power, battery-operated, mobile compute devices by everyday users, such volunteer-based computing projects are likely to suffer a significant setback in terms of the donated compute cycles. However, this challenge can be largely overcome if high-performing video-gaming hardware, which has significant longevity in the consumer market, is brought into the fold.

## III. Hardware Specifications

In this section, we document full details on the list of the video-gaming hardware devices we evaluated using the SHOC and the *Einstein@Home* benchmarks. We also include a high-end HPC GPGPU accelerator and a multi-core CPU for reference.

Basic hardware specifications for the devices tested are included in Table I. The Tegra X1 SoC, Radeon Fury GPU and the A10 APU are technologies commonly found in (or targeted at) consumer-grade gaming (high-end and mobile) devices, while the Tesla K40 GPGPU and x86 multi-core CPUs are utilized in HPC servers.

TABLE I
NVIDIA AND AMD HARDWARE SPECIFICATIONS

|  | Nvidia Tesla K40 | Nvidia Tegra X1 | AMD Radeon R9 Fury X | AMD A10-7850K | AMD FX-9590 |
|---|---|---|---|---|---|
| Device Type | GPU | SoC | GPU | SoC | CPU |
| Compute Cores | 2,880 | 256 | 4,096 | 512 | 8 |
| Base Clock (MHz) | 745 | 1,000 | 1,050 | 3,700 | 4,700 |
| Memory Clock (MHz) | 3,000 | 1,600 | 1,000 | 720 | 2,400 |
| Memory Bandwidth (GB/s) | 288 | 25.6 | 512 | 34.1 | 29.9 |
| Power (Watts) | 235 | 11 | 275 | 95 | 220 |
| GFLOPS | 4,290 | 512 | 8,600 | 856 | 150 |
| Launch Price | $5,499 | $599 | $649 | $173 | $269 |

The Nvidia Tesla K40 uses the Kepler GK110B GPU, a 7.1 billion-transistor chip manufactured by TSMC on a 28 nm process [27]. This discrete GPGPU was selected for its raw computational power and convenient actively cooled PCIe form factor. Nvidia specifically targets the HPC market with the Tesla series, particularly those segments that require ECC memory with the capability to correct single bit errors and detect double bit errors. However, enabling ECC results in a modest loss of memory capacity and performance. Molecular Dynamics simulations carried out on the XSEDE supercomputer *Keeneland* point to the questionable value of ECC memory for some areas of scientific computing [28].

The Nvidia Tegra X1 is a SoC device manufactured by TSMC on a 20 nm process and tested using the Jetson TX1 development board, the first of this class of devices to reach terascale performance [29]. Nvidia designed the Tegra processor with mobile devices in mind using quad core ARM Coretex A57 and A53 processors in a big.LITTLE design coupled with their latest power efficient Maxwell GM20B GPU. Nvidia is making a push into new markets with this device collaborating with numerous tablet and automotive manufactures. Note that OpenCL is currently not supported on the Tegra device. Therefore, we use Nvidia's own GPGPU framework (CUDA) to perform our tests.

The AMD Radeon R9 Fury X uses Fiji XT GPU that has 8.9 billion transistors and is manufactured by TSMC on a 28 nm process [30,31]. This enthusiast level discrete gaming GPU was the first to market with High Bandwidth Memory (HBM) that offers improvements in bandwidth and power efficiency compared to GDDR5 that is currently used in other GPUs.

The AMD A10-7850K APU uses a four core Steamroller based CPU and an eight core Radeon R7 Sea Islands based Graphics Core Next (GCN) GPU fused on the same die [32,33]. In total, the APU is a 2.41 billion transistor SoC manufactured by Global Foundries on a 28 nm process. The current generation video-gaming consoles (Sony's *PlayStation 4* and Microsoft's *Xbox One*) utilize this APU technology and intend to continue to do so in upcoming updates. This processor represents a big step forward for devices conforming to the newly instituted Heterogeneous System Architecture (HSA) where a system's main memory is shared between all compute cores rather then being segregated for CPU and GPU use [34].

The AMD FX-9590 is a traditional x86 CPU that features eight Piledriver Cores in the Vishera family [35]. This CPU was noteworthy in its capability to turbo clock up to 5 GHz while dissipating 220 Watts of heat. This CPU offers reasonable performance on single threaded applications, its computational performance places it well below the other devices tested, and is included in our study purely for the sake of comparison.

## IV. RESULTS AND DISCUSSION

In this section we present the detailed results of our extensive testing. In the appendix of this article, we include a table that enlists all the raw data generated by the SHOC benchmark suite. This may serve as a useful reference for readers interested in different types of applications. However, in the following section we focus on a select few benchmarks that are representative of the entire study.

### A.  Discrete GPU devices (Tesla K40 and Radeon Fury)

Nvidia first popularized the term GPU in 1990 before the release of their GeForce 256. This unit marked the beginning of a transition where 3D modeling processes such as graphical transform and lighting calculations were offloaded from the CPU to a discrete device. This proved advantageous for scientific computing as the processing power required to perform rapid large-scale calculations could be harnessed to run code containing parallel algorithms.

Consumer demands for increased realism in 3D games coupled with the development of low priced high-resolution displays has created a large and highly competitive market for GPU devices. Even today this demand continues to be pushed by 4K displays and virtual reality headsets.

The Nvidia K40 and AMD Fury X represent the highest performance discrete devices from both companies at the time of their release. Figure 1 below depicts the (single-precision floating-point) performance of these two devices in GFLOPS on a representative subset of SHOC benchmarks. The two devices offer remarkably similar performance despite substantial differences in their architecture, targeted market, and price points, showing the viability of gaming hardware for scientific computing. While close, the Fury X holds a slight lead over the K40 in all four single precision (SP) benchmarks as seen in Figure 1. The log-scale on the GFLOPS axis should be noted.

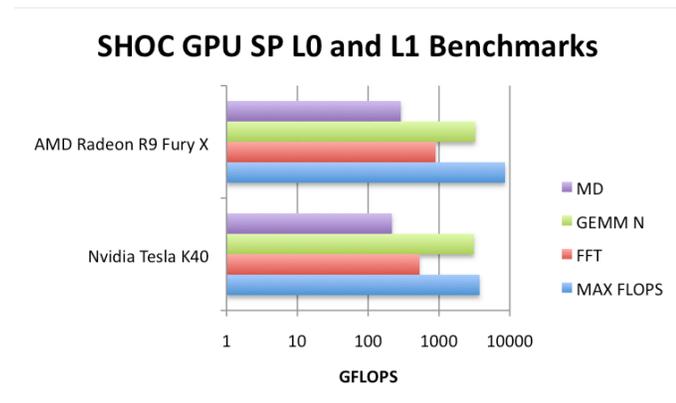

**Figure 1**

Since there are commonly expressed concerns on the double-precision (DP) floating-point performance of video-gaming devices, we also performed extensive testing in that context. In Figure 2 below, we show the same subset of SHOC benchmarks using double-precision data and operations instead. We note that the specialized HPC GPGPU Tesla K40 device does indeed perform better on most benchmarks, but its overall performance is on the same scale as the consumer-grade GPU Radeon device (whereas, its price is not).

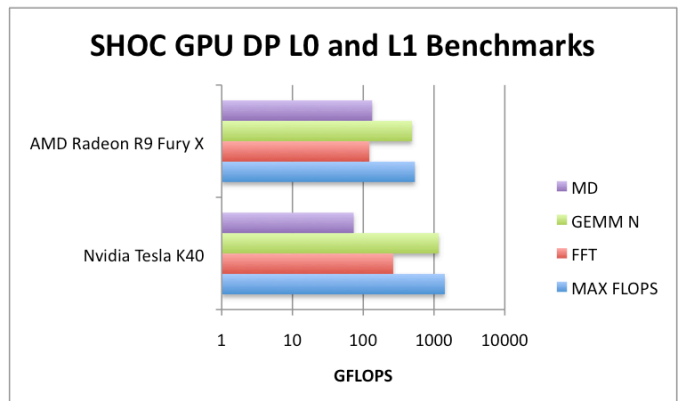

**Figure 2**

The benchmark results are in line with expectations as the K40 was designed with scientific computing in mind having a 1:3 DP to SP FLOP ratio. The gaming oriented Fury X has a 1:16 DP to SP FLOP ratio but still manages to come out ahead in the MD benchmark. The Nvidia GeForce 780 is a Kepler architecture based discrete gaming GPU with an introductory price matching the Fury X, but has 1/8th the DP to SP FLOP ratio of the K40 which would further erode its performance in DP benchmarks so was not included in the testing. Future generations of Nvidia GPUs are expected to further increase the gap in DP performance between the Tesla and GeForce lines.

One big drawback to a discrete card is the bottleneck imposed by PCIe bandwidth. Although each successive generation of PCIe has increased bandwidth, data transfer rates to-and-from the GPGPU are an order of magnitude lower than the system's internal main memory bandwidth. The limitations imposed by this low PCIe bandwidth are clearly seen in the SHOC L1 benchmarks shown in Figure 3, where each test is run with and without accounting for PCIe transfer time.

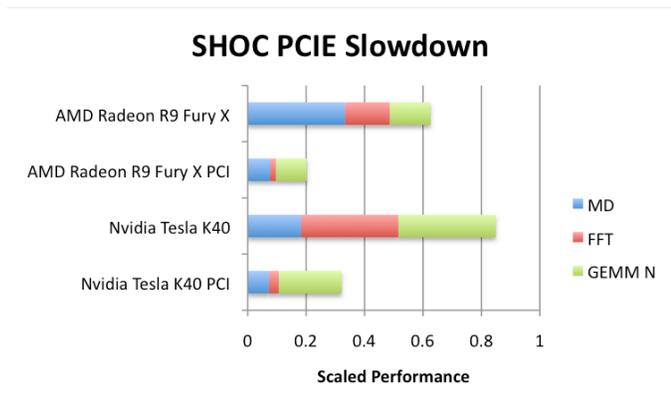

**Figure 3**

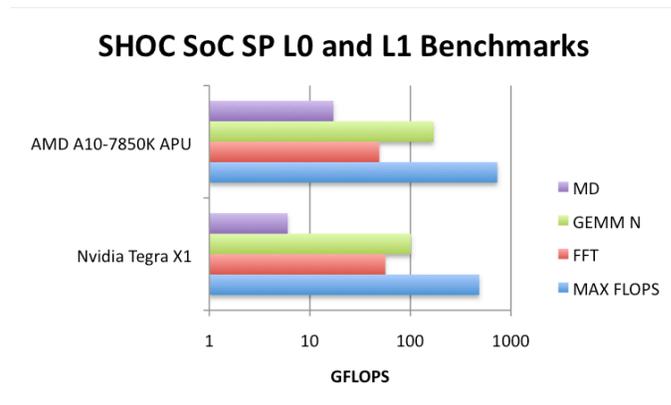

**Figure 4**

To emphasize the negative impact PCIe bandwidth can have, the results in Figure 3 are scaled to the best result for each test. Both MD and FFT show significant slowdown when PCIe bandwidth is taken into account, while GEMM N is less impacted.

Comparing the performance of the multi-core CPU on the same tests (data available in the appendix), one clear outcome is that GPU computing offers *order-of-magnitude* gains in overall performance over CPUs for a wide variety of scientific computing kernels. Moreover, this GPU-enabled acceleration can be achieved at very low cost, especially if the video-gaming class GPU that we mention in this section is utilized, as opposed to a specialized HPC GPGPU variant. There is a clear *order-of-magnitude* cost difference between these two devices, despite the performance being in the same ballpark.

### B. SoC and APU devices (Tegra X1 and APU A10)

Heterogeneous System Architecture (HSA) is a framework where more than one kind of processor or core can operate on the same tasks using a common bus with shared memory. This offers significant performance advantages and completely avoids PCIe bottlenecks, as a CPU and GPU can exist on the same substrate and calculations are performed in a unified memory space. It should be noted that we do not take advantage of this feature in this work; that requires OpenCL 2.0 which our current benchmark tests do not support.

APUs and SoCs are included in the HSA framework and are represented in this article by a "top down" design intended for a desktop computer in the AMD A10-7850K, that integrates a traditional x86-64 CPU and a GCN GPU. Also tested is a "bottom up" design intended for the tablet and mobile computing market in the Nvidia Tegra X1 featuring ARM cores in a big.LITTLE design and a Maxwell architecture based GPU.

Results for the single-precision SHOC SP L0 and L1 benchmarks are presented in Figure 4, with the A10-7850K and the X1 showing similar levels of performance despite being designed for very different markets.

While the A10-7850K leads slightly in the select benchmarks shown in Figure 4, the performance of the X1 is particularly impressive with a power consumption that is nine fold less than the A10-7850K! The double-precision SHOC DP L0 and L1 benchmarks follow a similar pattern with the A10-7850K performing particularly well on the GEMM benchmark as seen in Figure 5.

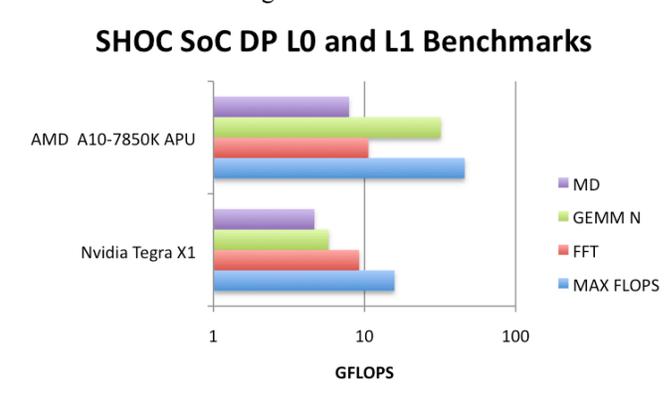

**Figure 5**

Once again, it is noteworthy that even the APU and SoC devices are powerful enough to deliver performance advantages in common scientific application kernels, as compared with multi-core CPUs. This is especially the case, when the power consumption of the devices is taken into account. The Tegra device's performance-per-Watt is two orders-of-magnitude higher than traditional CPUs. It is now common to assemble large clusters using thousands of compute devices for scaled up performance. Such clustered systems are severely limited by their significant power consumption and therefore, cooling needs. APU and SoC devices may offer interesting alternatives for power-efficient clustered supercomputing.

### C. Einstein@Home Performance

SHOC benchmarks may provide valuable insight into the computational power of a device, particularly if one's code mimics one of the included synthetic benchmarks. However, it is often useful to study the overall performance of a compute device on a complete end-to-end scientific application. In this section, we present such performance results using a single representative *Einstein@Home* binary radio pulsar (Parkes PMPS XT) work unit and present the results in Table II.

| Device | Time (kiloseconds) |
|---|---|
| AMD A10-7850K APU | 23.8 |
| AMD Radeon R9 Fury X | 2.30 |
| Nvidia Tegra X1 | 23.5 |
| Nvidia Tesla K40 | 3.46 |

The Fury X is particularly well suited for the OpenCL implementation of the *Einstein@Home* application with the K40 taking 1.5x as long to complete a work unit. The two SoC devices take nearly the same time to complete a work unit, but an order of magnitude longer than the discrete GPUs.

A particularly illuminating way to compare performance results on *Einstein@Home* are presented in Figure 6 where the actual energy necessary to complete a unit of work is found by multiplying the average completion time with the rate of power consumption (Watts) of the device.

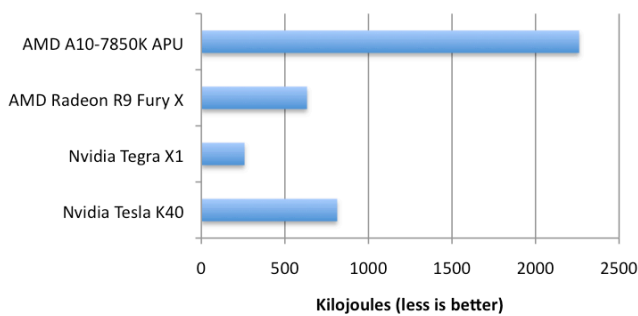

**Energy to Complete One Unit of E@H**

*Figure 6*

The tremendous effort made by Nvidia to improve performance per Watt is clearly apparent here with the Tegra X1 coming in 2.4x more efficient than the next closest device. This method of calculation does not account for the efficiency of power supplies or the power consumed by the CPU and components necessary to support the discrete GPUs. Those factors are highly dependent on the choice of components in a system and would only serve to increase the gap between the Tegra X1 and other devices.

While offering significant advantages in terms of performance per Watt, there is still a substantial difference in raw computational power with the Fury X completing 10x as many *Einstein@Home* work units as the Tegra X1 in the same period of time. This order-of-magnitude performance difference raises a host of additional issues including space and networking. A 4U server may hold as many as 8 GPUs while an equivalently powerful group of Tegra X1s may take up an entire rack of space. Network complexity would also see a substantial increase using Tegra X1s and potential upgrades to lower latency interconnects would be both limited and expensive. For this generation of SoCs, the best way to leverage their tremendous energy efficiency is to use them in a highly distributed "grid" like fashion such as the *Einstein@Home* project.

## V. Conclusions

In this article we present the results of our detailed evaluation of a variety of video-gaming hardware, repurposed for scientific supercomputing. We use efficiency metrics like the measured performance, performance-per-Watt and performance-per-dollar to compare the considered hardware devices with more traditional supercomputing hardware. The power efficiency metric, *performance-per-Watt* is considered vital currently because it is the key towards enabling the next generation of supercomputing.

We summarize the results of our study in the following list:

• *GPUs offer significant performance advantages*---Many-core GPUs typically perform an order-of-magnitude better over multi-core CPUs for a wide variety of scientific computational kernels. These benefits are present, not only in raw compute performance, but also when cost and power consumption is taken into account.

• *Video-gaming GPUs offer similar application acceleration to HPC GPGPUs at much lower cost*---Consumer-grade GPUs targeted at the video-gaming market offer similar performance to high-end, specialized HPC GPGPUs but at a small fraction of the cost. This is true even in the context of double-precision floating-point performance.

• *Mobile-device SoCs offer the best power-efficiency*---Unsurprisingly, technologies developed to operate on battery power (like phone and tablet hardware) has the best power-efficiency, by a significant margin. Harnessing this performance advantage for large-scale supercomputing is yet to be seen, although some exploratory projects have begun [26].

The authors would like to acknowledge support from the *Center for Scientific Computing and Visualization Research* at UMass Dartmouth. GK also acknowledges support from the National Science Foundation under award PHY-1414440 and PHY-1606333 and the Air Force Research Laboratory under CRADA agreement 10-RI-CRADA-09.

# VI. APPENDIX

TABLE III
## SHOC Benchmark Results

| Benchmark | Nvidia Tesla K40 | AMD Radeon R9 Fury X | Nvidia Tegra X1 | AMD Fusion A10-7850K APU | AMD FX-9590 CPU |
|---|---|---|---|---|---|
| bspeed_download (GB/s) | 10.5497 | 12.9545 | 10.4686 | 9.4583 | 22.5772 |
| bspeed_readback (GB/s) | 10.5584 | 14.1924 | 10.6373 | 10.0571 | 22.5452 |
| maxspflops (GFLOPS) | 3743.97 | 8563.42 | 484.522 | 733.97 | 77.5734 |
| maxdpflops (GFLOPS) | 1422.63 | 537.496 | 15.7751 | 45.9488 | 51.9592 |
| gmem_readbw (GB/s) | 177.498 | 487.63 | 15.1397 | 23.3226 | 0.6866 |
| gmem_readbw_strided (GB/s) | 18.2029 | 114.666 | 6.0576 | 15.3657 | 46.7156 |
| gmem_writebw (GB/s) | 173.663 | 447.993 | 12.7893 | 17.1887 | 0.3582 |
| gmem_writebw_strided (GB/s) | 7.2354 | 12.5633 | 1.9888 | 6.8982 | 43.4828 |
| lmem_readbw (GB/s) | 908.473 | 3436.8 | 215.476 | 266.708 | 61.6795 |
| lmem_writebw (GB/s) | 1136.65 | 3412.9 | 247.081 | 294.712 | 56.0024 |
| tex_readbw (GB/s) | 210.271 | 286.149 | 45.5643 | 63.3353 | 13.0108 |
| bfs (GB/s) | 1.2082 | 5.8897 | 0.2131 | 0.6879 | 0.195 |
| bfs_pcie (GB/s) | 1.0493 | 3.9657 | 0.1975 | 0.6375 | 0.1758 |
| bfs_teps (Edges/s) | 71060400 | 152044000 | 12752000 | 13731100 | 8027410 |
| fft_sp (GFLOPS) | 530.875 | 892.948 | 56.3586 | 49.0251 | 3.4051 |
| fft_sp_pcie (GFLOPS) | 53.1111 | 35.989 | 5.0751 | 17.1015 | 2.857 |
| ifft_sp (GFLOPS) | 530.544 | 812.385 | 56.4569 | 54.5175 | 3.4385 |
| ifft_sp_pcie (GFLOPS) | 53.2969 | 35.8458 | 5.3521 | 17.7244 | 2.8805 |
| fft_dp (GFLOPS) | 265.141 | 121.815 | 9.21 | 10.5905 | 2.5918 |
| fft_dp_pcie (GFLOPS) | 26.5918 | 16.156 | 2.1241 | 5.8503 | 2.009 |
| ifft_dp (GFLOPS) | 265.275 | 118.449 | 8.6632 | 10.3274 | 2.6343 |
| ifft_dp_pcie (GFLOPS) | 26.6117 | 16.0954 | 2.2504 | 5.7691 | 2.0344 |
| sgemm_n (GFLOPS) | 3115.49 | 3256.54 | 100.986 | 170.229 | 14.7719 |
| sgemm_t (GFLOPS) | 3127.91 | 774.361 | 93.6681 | 150.331 | 9.7866 |
| sgemm_n_pcie (GFLOPS) | 2170.4 | 2156.61 | 95.2578 | 156.642 | 14.6351 |
| sgemm_t_pcie (GFLOPS) | 2176.42 | 689.84 | 88.7196 | 140.126 | 9.7541 |
| dgemm_n (GFLOPS) | 1167.74 | 489.172 | 5.7714 | 31.9007 | 14.9615 |
| dgemm_t (GFLOPS) | 1234.17 | 494.417 | 5.849 | 33.6603 | 9.1538 |
| dgemm_n_pcie (GFLOPS) | 754.329 | 368.455 | 5.6417 | 30.0337 | 14.6656 |
| dgemm_t_pcie (GFLOPS) | 781.498 | 371.443 | 5.7158 | 31.4858 | 9.0292 |
| md_sp_flops (GFLOPS) | 216.703 | 288.933 | 6.0203 | 17.1391 | 5.0568 |
| md_sp_bw (GB/s) | 166.074 | 221.43 | 4.6137 | 13.1349 | 3.8754 |
| md_sp_flops_pcie (GFLOPS) | 41.4332 | 40.4597 | 2.7676 | 12.138 | 3.3798 |
| md_sp_bw_pcie (GB/s) | 31.7532 | 31.0071 | 2.121 | 9.3022 | 2.5902 |
| md_dp_flops (GFLOPS) | 73.277 | 133.729 | 4.6492 | 7.8838 | 4.24 |
| md_dp_bw (GB/s) | 98.3569 | 179.5 | 6.2405 | 10.5821 | 5.6911 |
| md_dp_flops_pcie (GFLOPS) | 29.3095 | 30.7099 | 2.4314 | 6.556 | 3.0611 |
| md_dp_bw_pcie (GB/s) | 39.341 | 41.2208 | 3.2636 | 8.7998 | 4.1088 |
| md5hash (GHash/s) | 2.5847 | 10.3697 | 0.5145 | 0.904 | 0.0947 |
| reduction (GB/s) | 159.242 | 161.513 | 16.3416 | 13.2883 | 0.5583 |
| reduction_pcie (GB/s) | 9.8453 | 11.7531 | 6.2787 | 5.4244 | 0.4694 |
| reduction_dp (GB/s) | 171.113 | 278.937 | 21.7324 | 12.9653 | 1.1003 |
| reduction_dp_pcie (GB/s) | 9.8886 | 12.0743 | 6.9257 | 5.3985 | 0.8218 |
| scan (GB/s) | 48.557 | 42.1121 | 6.5806 | 6.1895 | 0.0327 |
| scan_pcie (GB/s) | 4.7416 | 5.6862 | 1.7787 | 2.639 | 0.0321 |
| scan_dp (GB/s) | 43.2257 | 66.9224 | 6.1591 | 6.3924 | 0.0654 |
| scan_dp_pcie (GB/s) | 4.6834 | 5.9863 | 2.2345 | 2.6887 | 0.0635 |
| sort (GB/s) | 3.06 | 0.7462 | 0.4447 | 0.2309 | 0.0003 |
| sort_pcie (GB/s) | 1.936 | 0.666 | 0.4084 | 0.2199 | 0.0003 |
| spmv_csr_scalar_sp (GFLOPS) | 2.4467 | 4.0236 | 0.6716 | 0.3282 | 1.0817 |
| spmv_csr_scalar_sp_pcie (GFLOPS) | 1.2389 | 1.2893 | 0.1879 | 0.2838 | 0.3013 |
| spmv_csr_scalar_pad_sp (GFLOPS) | 2.9054 | 4.4426 | 0.7561 | 0.3101 | 1.0843 |
| spmv_csr_scalar_pad_sp_pcie (GFLOPS) | 1.3572 | 1.8123 | 0.5145 | 0.2725 | 0.2592 |
| spmv_csr_vector_sp (GFLOPS) | 18.4959 | 25.8192 | 2.9135 | 2.8263 | 0.9701 |
| spmv_csr_vector_sp_pcie (GFLOPS) | 2.2118 | 1.7669 | 0.2397 | 1.2039 | 0.292 |
| spmv_csr_vector_pad_sp (GFLOPS) | 20.2005 | 27.2547 | 2.9742 | 2.9978 | 0.9709 |
| spmv_csr_vector_pad_sp_pcie (GFLOPS) | 2.2613 | 2.7515 | 1.0407 | 1.285 | 0.2521 |
| spmv_ellpackr_sp (GFLOPS) | 17.5968 | 11.1322 | 2.6979 | 3.7735 | 0.454 |
| stencil (GFLOPS) | 128.805 | 354.875 | 0.3184 | 14.453 | 0.5359 |
| stencil_dp (GFLOPS) | 57.6011 | 148.964 | 0.109 | 9.4046 | 0.5063 |
| triad_bw (GB/s) | 13.6654 | 10.9797 | 6.6658 | 5.8931 | 3.1222 |
| s3d (GFLOPS) | 97.9084 | 121.407 | 7.2195 | 8.0729 | 0.8412 |
| s3d_pcie (GFLOPS) | 83.3313 | 97.3392 | 6.7368 | 7.9218 | 0.8336 |
| s3d_dp (GFLOPS) | 51.411 | 48.8817 | 4.2222 | 3.6208 | 1.3738 |
| s3d_dp_pcie (GFLOPS) | 43.4497 | 41.2434 | 4.1375 | 3.5417 | 1.354 |